\documentclass[10pt, conference, compsocconf]{IEEEtran}
\usepackage{graphicx}
\usepackage{fancyhdr}
\DeclareGraphicsExtensions{jpg}
\ifCLASSINFOpdf
\else
\fi

\begin{document}
\title{SDFog: A Software Defined Computing Architecture for QoS Aware Service Orchestration over Edge Devices}

\author{\IEEEauthorblockN{Harshit Gupta}
\IEEEauthorblockA{School of Computer Science\\
Georgia Institute of Technology\\
Atlanta, US\\
Email: harshitg@gatech.edu}
\and
\IEEEauthorblockN{Shubha Brata Nath, Sandip Chakraborty, Soumya Kanti Ghosh}
\IEEEauthorblockA{Department of Computer Science and Engineering\\
Indian Institute of Technology Kharagpur\\
Kharagpur, India\\
Email: nath.shubha@gmail.com, sandipc@cse.iitkgp.ernet.in, skg@iitkgp.ac.in}
}

\maketitle

\thispagestyle{fancy}

\begin{abstract}
Cloud computing revolutionized the information technology (IT) industry by offering dynamic and infinite scaling, on-demand resources and utility-oriented usage. However, recent changes in user traffic and requirements have exposed the shortcomings of cloud computing, particularly the inability to deliver real-time responses and handle massive surge in data volumes.  Fog computing, that brings back partial computation load from the cloud to the edge devices, is envisioned to be the next big change in computing, and has the potential to address these challenges. Being a highly distributed, loosely coupled and still in the emerging phase, standardization, quality-of-service management and dynamic adaptability are the key challenges faced by fog computing research fraternity today. This article aims to address these issues by proposing a service-oriented middleware that leverages the convergence of cloud and fog computing along with software defined networking (SDN) and network function virtualization (NFV) to achieve the aforementioned goals. The proposed system, called “Software Defined Fog” (SDFog), abstracts connected entities as services and allows applications to orchestrate these services with end-to-end QoS requirements. A use-case showing the necessity of such a middleware has been presented to show the efficacy of the SDN-based QoS control over the Fog. This article aims at developing an integrated system to realize the software-defined control over fog infrastructure.
\end{abstract}


%
\IEEEpeerreviewmaketitle

\section{Introduction}
The paradigm of computing has seen an evolution that has revolved around user requirements, starting from mainframe computing, through personal computers, into cloud computing.  Computing in the cloud offers advantages like ubiquitous access, infinite scalability and utility-oriented usage, and has been called the next big thing in IT. However, recently emerging applications - demanding real-time response or generating voluminous data - expose the shortcomings of cloud computing, which being centralized and remotely located does not perform well with these requirements. With the ever increasing number of connected endpoints (estimated at 50 billion by 2020 \cite{cisco_iot}), the global mobile traffic is expected to increase from 2.6 to 15.8 exabytes by 2018 \cite{abdelwahab2016network}, a surge that demands scalability which the cloud paradigm lacks.
To address these issues, the computing world is taking measures to distribute computation, so that datacenters are not the only the places where applications are hosted. Compute, storage and network services should be available towards the edge devices. This change of philosophy can be realized by a continuum of resources from the network edge, through the core network, to the datacenters - what has been rightly termed Fog Computing.
\par Fog computing, term coined by Cisco \cite{bonomi2012fog} ,refers to extending cloud computing to the edge of an enterprise's network. Fog computing facilitates the operation of compute, storage and networking services between end devices and cloud computing datacenters                                     \cite{webopedia}. Such a large number of devices involved make application provisioning and infrastructure configuration and management complicated.  The concept of SDN \cite{nunes2014survey, kim2013improving} has emerged as a popular tool for the management of network services through abstraction of low-level functionality on a large cluster of network devices. We envision that concepts of SDN can be applied to fog computing, at a more abstract level, allowing centralized control over computation, storage and network resources in fog infrastructure. Contemporary SDN systems are designed to answer the question of 'Where to send received packets?', where a “Software Defined Fog” (SDFog) will be able to answer the broader and more pertinent question 'What to do with received packets?', whether to process it by an application at the receiving machine, or forward it to another one. In this paper, we give substance to this vision of software-defined control in the fog by proposing a mechanism for orchestrating services hosted across the network in a QoS aware fashion. 

\begin{figure}[]
\centering
\includegraphics[width=0.5\textwidth]{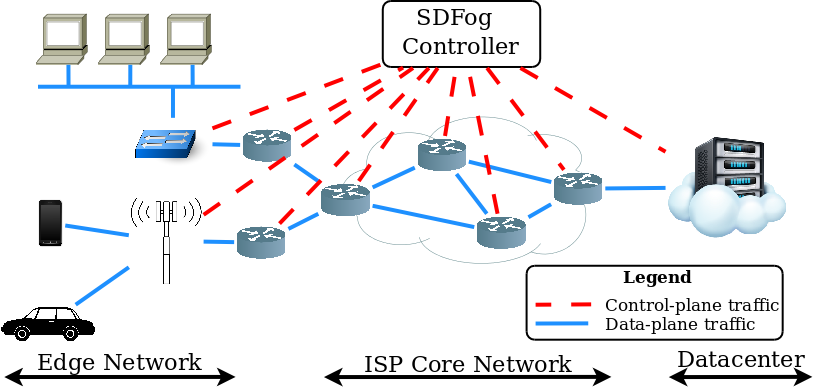}
\caption{High level overview of Software Defined Fog environment. The SDFog Controller is logically centralized and may be implemented as a multi-domain controller}
\label{fig:overview}
\end{figure}

\par This article proposes a service-oriented middleware that distributes service hosting throughout the fog environment, making all the nodes in the resource continuum - from the edge to the cloud - capable of hosting services. The proposed system performs QoS aware orchestration by scheduling flows between services that satisfy specified QoS constraints. QoS preserving virtual network functions (VNFs) \cite{han2015network} are used for establishing healthy user experience, which are allowed to run on the same infrastructure used by user applications and access resources of the underlying devices through a hypervisor. The aim of the article is to extend the concept of SDN to the application layer, and into a "\textit{Software Defined Fog}" (SDFog), which is able to execute data-plane functions dealing with compute, storage and network resources on fog nodes. We demonstrate the uses of SDFog architecture with the help of a use case from smart home paradigm, and analyze the effectiveness of this architecture through a small-scale prototype implementation. 

\section{Need for convergence of SDN and SOA in Fog Computing}
In this section, we first discuss the requirements to develop a converged architecture by considering the concepts of SDN and SOA. Based on these requirements, we shall progress to develop the overall architecture of SDFog that can cater the present requirements.  
\subsection{Service Orientation in the IoT}
The proliferation of the IoT has brought about a thrust in the Fog computing  paradigm, and is one of the most important and pertinent use-cases of the Fog. IoT poses some unique requirements for the infrastructure to meet, which have been listed below.
\begin{itemize}
\item \textbf{Enormous scale }: IoT services are essentially provided by physical devices connected to the Internet. Typically, data generated and events raised by these real-world objects are available globally - and need to be orchestrated for accomplishing complex tasks. The number of such devices connected to the Internet has been foresaid to be orders of magnitude larger than the number of computers connected to the Internet today \cite{mietz2013semantic}. Such an increase in the number of potential service providers requires the discovery mechanism to be scalable with respect to the number of objects.
\item \textbf{Heterogeneity of devices }: Devices in the IoT, coming from a variety of vendors, are widely heterogeneous, with highly varying sensing/actuating characteristics and communication protocols. Such heterogeneity demands an adapter-based approach that provides uniform and seamless access mechanisms. Furthermore, abstracting physical devices into high-level entities with associated metadata information about device parameters is necessary to perform pertinent orchestration in such a heterogeneous environment.
\item \textbf{Unknown service availability }: The topology of the IoT is quite dynamic and unknown, making availability of components during orchestration dynamic and unpredictable. Typing up such dynamic entities for accomplishing a task requires abstracting them in the form of composable services with associated metadata describing it, so that they can be discovered and orchestrated on the fly.
\item \textbf{Locality-preserving }: The property of content and path locality of service discovery and orchestration is very important for IoTs \cite{li2015decentralized}. Content locality improves security by placing the metadata of services in the same administrative domain as the device realizing the service. Path locality guarantees that a routing message between two entities within an administrative domain always stays within that domain’s boundary. This isolation is instrumental in avoiding malicious  attacks by foreign machines outside the administrative domain. Storing metadata on and orchestrating services through the cloud does not follow these principles and hence exposes the system to security risks. 
\end{itemize}
The proposed system addresses these issues by being distributed over all fog nodes - thus providing scalability, providing an adapter-based approach to support heterogeneous service providers, discovering services using dynamically and preserving the content and path locality property by hosting services on premise. Accordingly we analyse next, why SDN can play an important role to such service orchestration. \subsection{SDN for QoE management}
Requirements of the present day user traffic have gained heterogeneity, with latency-critical traffic sharing the same infrastructure as high-bandwidth delay-insensitive ones. Such heterogeneous requirements pose the need of handling disparate traffic according to different policies, which can be dictated by the applications to which the traffic belongs. Such a kind of network management provides an equally good experience to all users by allocating unequal resources to their flows, rather than provisioning equal resources to all user flows and jeopardising the experience of users with stricter requirements.
SDN involves the separation of the control and data plane, allowing the network to be programmed and controlled centrally and has enabled the abstraction of network parameter specification into high-level policies. SDN has provided the means to develop network parameter control policies that can be aware of dynamic QoE of users, and has emerged at a very appropriate time - when mobility and heterogeneity have become important challenges in network management. 
Research probes in this direction have shown that it is indeed possible to meet QoS constraints by dynamically configuring network devices carrying the traffic via software defined networking \cite{akella2014quality, egilmez2012openqos}. Optimal routing and bandwidth allocation have been the fundamental methodologies of existing QoS-aware flow management. We, however, envision that the state-of-the-art in providing end-to-end QoS guarantees can be pushed further by instantiating QoE enhancing VNFs at appropriate points in the network – where standard techniques fail to provide the required service guarantees. \cite{alaska} proposes a VNF that reduces bandwidth consumption and latency of operations by data-deduplication and TCP optimizations, which is able to garner 60-95\% reduction in wide area network (WAN) utilization. The optimal placement of network functions like application acceleration often is  upstream of access paths - on customer premises \cite{dist_nfv}. Fog computing offers a dynamic and elastic method of provisioning VNFs by virtualizing the resources on network devices between the edge and cloud. Next we discuss the details architectural view of SDFog paradigm.

\section{Architecture}
The architecture focuses on design of a fog node (shown in Figure 1), where two major components are the “Service Oriented Middleware” (SOM) and the “Distributed Service Orchestration Engine” (DSOE). These components run on all fog nodes in the physical infrastructure, interfacing with and utilizing services provided by the operating system and hypervisor on the devices. SOM abstracts device functionalities and data sources and exposes them as services on a fog node, which can, at a low level, be provided by dedicated hardware connected to it (through device drivers) or by application modules running on the fog node. DSOE performs the task of discovering, deploying and orchestrating services and performing QoS-preserving operations.
\begin{figure}[]
\centering
\includegraphics[width=0.5\textwidth]{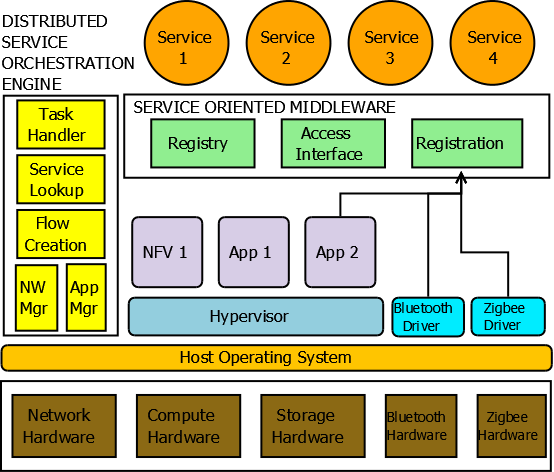}
\caption{Architecture of a fog node showing the proposed Middleware and Orchestration Engine. User applications and VNFs run on the same platform through hypervisor.}
\label{fig:arch}
\end{figure}
\subsection{Service Oriented Middleware}
The middleware component performs the task of managing and providing interface to services on a fog node. The component is capable of existing independently on a fog node - dealing with services hosted on that node only, however interaction with the DSOE is required for orchestration of services hosted at different nodes in the network.
Services can be hosted on a fog node by 
\begin{itemize}
\item applications running on the fog node utilizing its resources through the hypervisor, for example, a smart home control service (accessed by smart home smartphone application) hosted on the home gateway (by smart home gateway application), or
\item devices connected to the fog node through device-specific hardware, for example, a Bluetooth light-sensing device, connected to a fog node, hosts it sensing service via the Bluetooth driver.
\end{itemize}
 
The service-oriented middleware allows every fog node in the resource continuum between the network edge and core to act as a service hosting platform and performs the following functions.

\subsubsection{Service Registration}
Service registration can be carried out by the entity hosting the service (fog application or dedicated device hardware) by sending a service-registration message to the hosting fog node, accompanied by associated metadata which contains information about the service, e.g. parameters of the service-providing device, and will be used when orchestrating multiple services under QoS constraints.\subsubsection{Access Interface}
The middleware provides an interface to applications running on a fog-node to access the services hosted on it. The interface can be implemented via a web service or using publish-subscribe protocols. This interface would allow application modules to access data or to invoke functionalities offered by the service-providing entity.\subsubsection{Registry}
Service registry allows a requester to discover services hosted on a fog node. Associated with each service is its metadata - which allows service selection under constraints.

\subsection{Distributed Service Orchestration Engine}
The DSOE glues together instances of SOM running on fog nodes throughout the network. Applications specify high-level tasks as a directed acyclic service graph, in which edges (denoting data flows between services) annotated with the required QoE parameters. The accomplishment of such a high-level task requires the following sub-tasks to be carried out:
\begin{enumerate}
\item Decomposition of high-level task into component services 
\item Discovery of component services 
\item Calculation of network parameters for achieving specified QoE along all flows between the chosen services
\item Application of calculated network parameters on all devices carrying data-plane traffic for the flows
\end{enumerate}
These individual tasks are carried out by specific modules in the distributed service orchestration engine, as illustrated in Figure 2.

\begin{figure}[]
\centering
\includegraphics[width=0.5\textwidth]{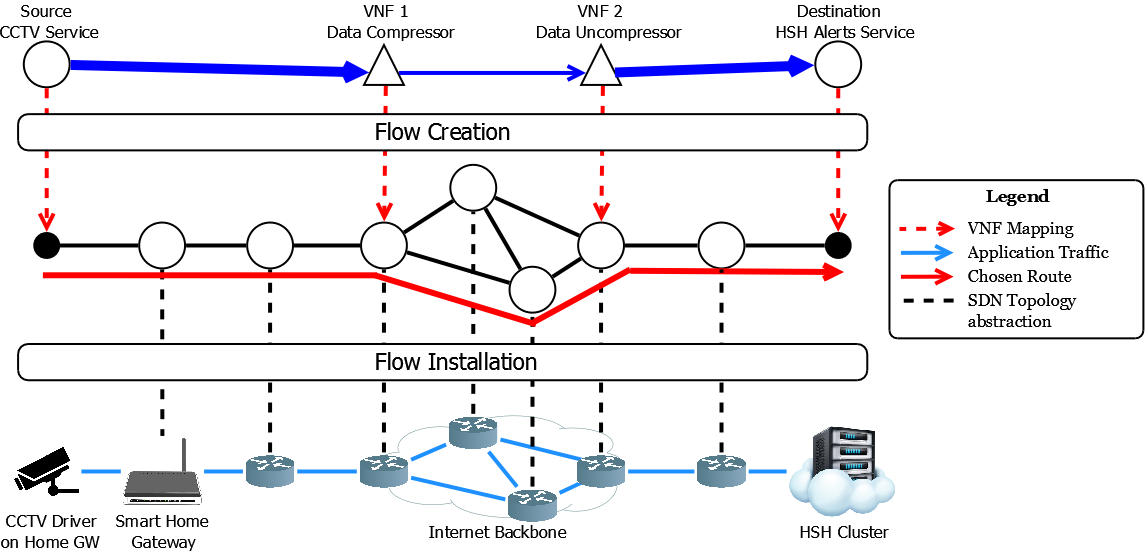}
\caption{Schematic representation of steps involved in task deployment following service discovery. Flow creation computes flow path based on current state of network topology. Flow installation sends VNF instantiation and flow table update commands to concerned fog nodes.}
\label{fig:mappings}
\end{figure}

\subsubsection{Service Discovery}
Submission of a high-level task is followed by the discovery of component services which need to be orchestrated. Discovery of service can be carried out in a distributed manner – with the "\textit{Service Discovery}" module on requesting fog node making a recursive query to its north and south neighbour fog nodes. Service discovery can also be done by utilizing the knowledge of global topology possessed by the SDFog Controller – by sending the discovery request to the controller. \par In the event when no instance of the requested service is found, a request for service instantiation is sent to the SDFog Controller. If the service can be provided by a software application, the controller determines the point-of-deployment and sends application instantiation command to the node where the service should be deployed. This command is received by the "\textit{App Manager}" module on the chosen fog node, which creates and deploys an instance of the corresponding application on the device's hypervisor - which in turn hosts the service by registering itself on the node's middleware. Upon successful discovery/creation, details of the requested services are sent back to the requesting application.

\subsubsection{Flow Creation}
Orchestration of services essentially means flow of data between them, and often such flows have associated constraints on network performance parameters like delay, jitter, packet loss, etc. Flow creation builds an overlay network of orchestrated services over the physical network topology in a QoS-satisfying fashion. Discovery of appropriate services for a task allows the DSOE to know network parameters like data rate, point-of-deployment, etc. which will be instrumental in determining low-level requirements of data flows, like bandwidth consumption, latency etc. helping the DSOE in choosing forwarding paths for the service overlay network. The complex and heterogeneous network between the point-of-deployment of orchestrated services calls for identifying bottlenecks in the forwarding path and taking measures for mitigating the effect of such a bottleneck, either by shaping traffic or changing the forwarding path.
\par Network function virtualization \cite{han2015network, jain2013network} allows flexible placement of network functionality at different points in the network topology, which can be leveraged to deploy QoE-enhancing functions like traffic shapers or WAN accelerators at points where congestion can deteriorate user experience. 
\par Flow creation, hence, involves (a) instantiation of appropriate VNFs in the network at appropriate places (b) installation of necessary forwarding rules to create a path between services. Flow creation for assuring QoS has been a well studied problem in the context of WANs \cite{seddiki2014flowqos, hu2015sdn}. These works throw light on the role of both VNFs and forwarding rules for QoS management. 
\subsubsection{Flow Installation}
Flow installation involves sending control information/commands to network devices situated in the path chosen for the flow. Control information comprises flow table entries as well as VNF instantiation commands, in order to setup an end-to-end flow that meets the QoS requirements of the application.
\par Flow installation commands are sent from the SDFog \textit{Controller} to \textit{Network Manager} and \textit{App Manager} components of the Distributed Service Orchestration Engine on nodes in physical network infrastructure. The \textit{App Manager} handles the  deployment of relevant VNFs using services provided by the device's hypervisor. The Network manager, on the other hand, is responsible for applying flow table entries on underlying network forwarding hardware for overlaying the forwarding path decided by the \textit{Flow Creation} module.
\par Upon application of flow installation commands by the receiving fog nodes, data flowing from source to destination service begins following the chosen path and adheres to QoS constraints dictated by the application.

\section{Health Smart Home : A motivating use-case}
\begin{figure}[]
\centering
\includegraphics[width=0.5\textwidth]{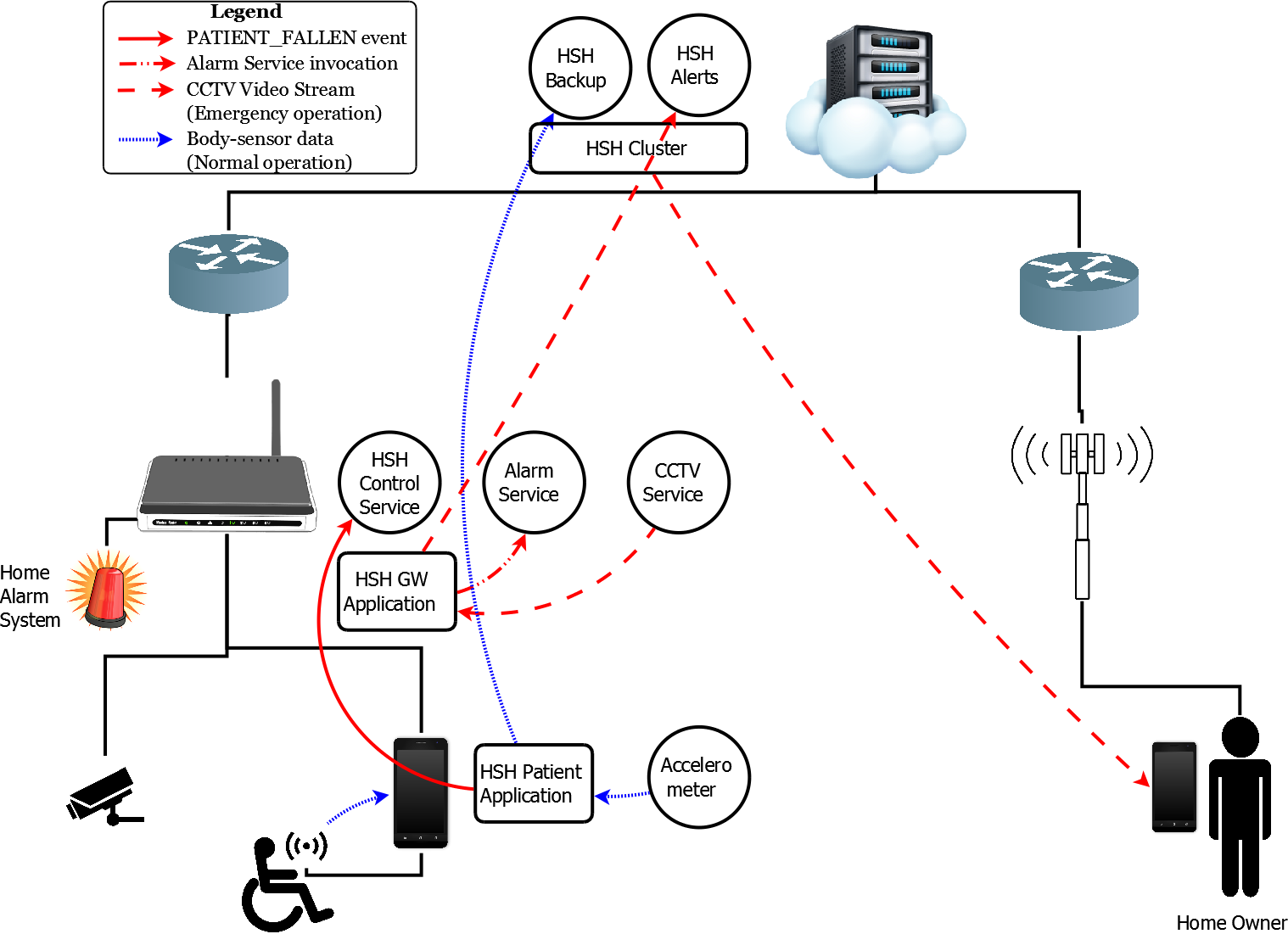}
\caption{Illustration of use-case Health Smart Home. Both Normal and Emergency modes of operation and the services involved in these modes are shown.}
\label{fig:usecase}
\end{figure}

Smart Homes have garnered a lot of popularity in recent times and offer some of the typical use-cases of IoT. The multitude and heterogeneity of devices in such a setting make their context-aware orchestration challenging, and this section shows how the proposed middleware can facilitate that. Healthcare applications form an exemplary candidate for context-aware computing, wherein the workflow of an application is sensitive to the context sensed by it. This section details a “\textit{Health Smart Home (HSH)}” application as a use-case for the proposed architecture and brings out the need for such an orchestration platform. Figure \ref{fig:usecase} illustrates the use-case with the constituent services that need to be orchestrated in order to fulfil the requirements of the HSH application.

\par The connectivity of the use-case can be described in the following points :
\begin{enumerate} 
\item A wheelchair-bound patient has wearable body area network sensors able to measure physical properties like acceleration, heart-rate, etc. which are connected to the patient's smartphone via a wireless protocol (e.g. Zigbee, Bluetooth, etc.). These sensors are able to register themselves as a data-providing service on smartphone's service-oriented middleware (through Zigbee or Bluetooth device drivers), thus making sensed data available to requesters of the service. 
\item Patient's smartphone runs an instance of HSH patient application, which subscribes to accelerometer service in order to receive real-time measurements of patient's acceleration. 
\item Smart home devices like CCTV Cameras and Alarm Systems are registered on Home Gateway fog node as services which can be invoked to, say, retrieve real-time video streams or trigger an alarm respectively. The “\textit{Home Gateway}” runs an instance of HSH gateway application and makes HSH control service available for the smartphone application counterpart to connect to it.
\item The cloud datacenter hosts virtual machines running HSH cluster application – which make HSH backup and HSH alert services available for coordination and storage purposes.
\end{enumerate}
The system has two operational modes, \textbf{normal} and \textbf{emergency}, wherein emergency operation mode is triggered by sensing appropriate context of the patient. These modes are described in the following :
\begin{enumerate}
\item \textbf{Normal operation} : The “\textit{HSH Patient Application}” aggregates information from body area network services, compresses it and sends it to \textit{HSH Cloud Backup Service} for storage and later analytics. In addition, it continuously mines the stream of acceleration values (from the accelerometer service) to detect events like falling of patient, in which case the \textit{HSH Control Service} is alerted and emergency operation is triggered.
\item \textbf{Emergency operation }: Upon receipt of an emergency event, \textit{HSH Gateway Application} triggers \textit{Home Alarm System} by invoking the service exposed by it on \textit{Home Gateway}. The application sets up real-time live CCTV footage streaming to the owner's smartphone. In order to do so, \textit{HSH Gateway application} invokes a task orchestrating the CCTV service on the \textit{Home Gateway}, and the \textit{HSH Alerts Service} running on a cloud datacenter (home owner is connected to \textit{HSH Alerts Service} for receiving emergency notifications). The \textit{Alerts Service} then handles the streaming of video from the datacenter to home owner's smartphone.
\end{enumerate}

\begin{figure}[]
\centering
\includegraphics[width=0.5\textwidth]{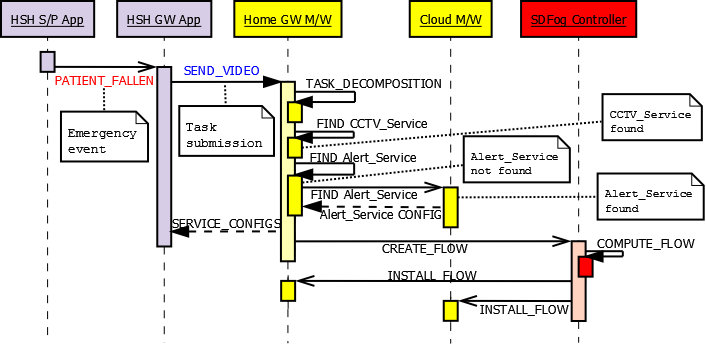}
\caption{Sequence diagram for Emergency operation mode in HSH use-case. Distributed service discovery through communication between instances on different fog nodes is shown.}
\label{fig:sequence}
\end{figure}

The invocation of the emergency operation mode involves functions of all components in the DSOE. Figure \ref{fig:sequence} explains the workflow of the system in this scenario. The event of the patient falling is detected by the \textit{HSH smartphone application} by listening to and mining data stream from \textit{Accelerometer service} registered on the smartphone. On detection of this event, it informs the \textit{HSH Control Service}, which submits the task of sending video stream to the \textit{HSH Alert Service} to the gateway's middleware. The middleware decomposes the received task into the component services and discovers them in a distributed manner - \textit{CCTV service} is found on the gateway itself, while \textit{Alerts Service} is found hosted on the \textit{HSH Cloud cluster}. Information about these services are sent to the requesting application - which schedules data flow between them. The middleware on the gateway then submits this flow and associated device parameters and QoS constraints to \textit{SDFog controller} for flow creation and installation. The controller computes a flow using its knowledge of the network topology and sends instructions to the involved physical devices to install the created flow.
\subsection{Evaluation}
For the purpose of demonstrating benefits of deploying VNFs in WAN for QoS management, an experiment emulating the emergency HSH operation was conducted over the \textit{Mininet emulator} platform that emulates a real scenario by creating virtual instances of network devices (using \textit{network namespace} and \textit{openvSwitch}). We have created a small scale emulation topology by creating virtual instances of network devices, following the topology architecture giving in Figure \ref{fig:usecase}. The network devices are implemented over a virtual machine with a network namespace. The emulated scenario has the \textit{HSH Control service}, sending real-time stream from a CCTV camera in the smart home to the \textit{HSH Alerts Service}, which in turn streamed the video to the home owner. In addition to this flow, the scenario has other users connected to the Internet performing activities like streaming videos leading to a high traffic in the backbone network, simulated by 250 kbps flows from streaming server to user's device. The backbone has been abstracted and emulated by a 1 Mbps link, with 20 millisecond latency from edge to core. The experiment compares the video data arrival curves in case of a QoS aware flow creation approach against the traditional best-effort routing approach. \textit{Mininet emulator} is used for topology emulation and POX is used as the SDN controller, while the captured video is streamed using the UDP streaming protocol.

\par Based on device parameters contained in metadata of CCTV service, the \textit{HSH Alerts service} requests for reserving 500 KB end-to-end from HSH cluster to the HSH owner's smartphone. In case of QoS-aware routing, a bandwidth-reservation VNF is placed in the backbone network that guarantees sufficient bandwidth for the emergency video stream.
\par
\begin{figure}[]
\centering
\includegraphics[width=0.5\textwidth]{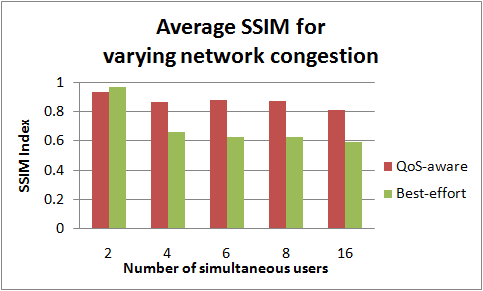}
\caption{Average SSIM analysis}
\label{fig:avg_ssim}
\end{figure}

 In order to compare the performance of QoS-aware against best-effort routing policy, we use the video stream's per-frame Structural Similarity Index (SSIM) \cite{wang2004image} to source video. Figure \ref{fig:avg_ssim} compares the average per-frame SSIM of video received by HSH Owner in QoS-aware and best-effort routing, with varying number of users generating traffic - providing varying levels network congestion.

\begin{figure}[]
\centering
\includegraphics[width=0.5\textwidth]{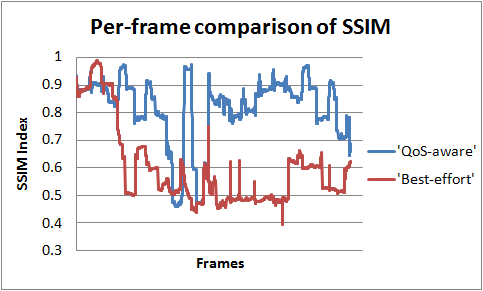}
\caption{Per frame SSIM comparison}
\label{fig:ssim}
\end{figure}

Figure \ref{fig:ssim} shows one of the cases when there are 16 other users running applications in the background, and compares the SSIM of each frame to the source video. The results from the figure show that with increasing network congestion levels, the QoS-aware approach is able to deliver a steady average video quality, while the quality of video delivered by best-effort approach falls with increasing network congestion. 
Based on the results shown, guaranteeing QoS to users by instantiating VNFs is possible and functions implementing network level optimizations or data deduplication can be used to satisfy other sorts of QoS requirements like latency or throughput respectively. This small scale prototype implementation provides a proof-of-concept for the applicability and benefits of using the SDFog paradigm that can utilize the computation capability of the edge devices dynamically by using the SDN based controller.

\par The results indicate that QoS-aware routing leads to a better rate of data delivery to the Health Smart Home owner, thereby providing better QoE. Besides, due to heavy network congestion, the TCP connection in best-effort routing is terminated before the entire video is streamed. 

\section{Open Challenges and Future Research Directions}
Realizing service-orientation in fog computing is fraught with a number of issues, particularly because of the widely distributed nature of the infrastructure. Incorporating semantics for service orchestration in the fog is a major challenge, and one which can solve issues like service conflicts, wherein multiple applications use the same actuator service and send conflicting actions, and the actuator service should decide - with the help of a knowledge base - the final action to be taken so that the QoS of all requesting services are not heavily jeopardised. Enforcing semantics in a distributed environment has been studied in the past \cite{mietz2013semantic, el2013semantic} and these studies can be used for doing so in the fog.
\par One of the key challenges is the extension of the south-bound SDN API with the help of NFV \cite{sezer2013we} to incorporate information governing all resources  - compute, storage and network - as opposed to the current APIs that deal only with network information. This would allow seamless extension of the current SDN architecture for controlling the fog infrastructure.

\par Making network configuration as simple and high-level as possible has been the goal of research in the SDN domain, which involves the development of appropriate north-bound interfaces connecting the \textit{SDFog Controller} with the applications submitting QoS annotated tasks. Applications should be able to specify required QoS parameters for a given task at an abstract level, which should then be translated into low-level network decisions by the SDFog controller.

\par Discovery of fog computing resources in the user's proximity and deployment of appropriate application components so as to satisfy the application's QoS requirements form one of the key milestones towards seamless user experience. Saurez et al. \cite{saurez2016incremental} have proposed an incremental application deployment and migration mechanism that makes use of a dynamic discovery and deployment protocol to achieve this. 
\par A significantly large fraction of contemporary devices are mobile and mobility, hence, forms one of the key challenges to be addressed. Application containerization and efficient state migration approaches are required so that user experience is not affected in the event of a change in the connection access point. Such issues can be addressed with the help of software-defined control wherein policies for migrating user-state can be initiated on detection of user's connection to another access point.

\section{Conclusions}
In this paper the design of a service-oriented approach for orchestrating devices in a Fog computing environment has been proposed, that abstracts every entity as a service with associated metadata. The system provides an interface to applications for submitting service orchestration tasks which can be bound by QoS constraints. Distributed service discovery is performed, followed by QoS aware flow creation and installation by the SDFog controller. This article proposes the use of virtual network functions for improving QoS parameters and describes an exemplary use-case where the user experience can be improved by installing VNFs at appropriate points in the network. 
\par This work tends to amalgamate the independently researched areas of SDN, NFV and fog computing to enable the seamless management of resources and orchestration of functionalities in a heterogeneous and widely distributed environment.

\bibliographystyle{plain}
\bibliography{references}

\begin{thebibliography}{10}

\bibitem{abdelwahab2016network}
Sherif Abdelwahab, Bechir Hamdaoui, Mohsen Guizani, and Taieb Znati.
\newblock Network function virtualization in 5g.
\newblock {\em IEEE Communications Magazine}, 54(4):84--91, 2016.

\bibitem{akella2014quality}
Anand~V Akella and Kaiqi Xiong.
\newblock Quality of service (qos)-guaranteed network resource allocation via
  software defined networking (sdn).
\newblock In {\em Dependable, Autonomic and Secure Computing (DASC), 2014 IEEE
  12th International Conference on}, pages 7--13. IEEE, 2014.

\bibitem{bonomi2012fog}
Flavio Bonomi, Rodolfo Milito, Jiang Zhu, and Sateesh Addepalli.
\newblock Fog computing and its role in the internet of things.
\newblock In {\em Proceedings of the first edition of the MCC workshop on
  Mobile cloud computing}, pages 13--16. ACM, 2012.

\bibitem{cisco_iot}
Cisco.
\newblock New cisco internet of things (iot) system provides a foundation for
  the transformation of industries.

\bibitem{alaska}
Alaska Communications.
\newblock How does wan optimization work?
\newblock https://www.youtube.com/watch?v=6mAP70cO4WM.

\bibitem{dist_nfv}
CIMI Corp.
\newblock Is 'distributed nfv' teaching us something?
\newblock http://blog.cimicorp.com/?p=1453.

\bibitem{egilmez2012openqos}
Hilmi~E Egilmez, S~Tahsin Dane, K~Tolga Bagci, and A~Murat Tekalp.
\newblock Openqos: An openflow controller design for multimedia delivery with
  end-to-end quality of service over software-defined networks.
\newblock In {\em Signal \& Information Processing Association Annual Summit
  and Conference (APSIPA ASC), 2012 Asia-Pacific}, pages 1--8. IEEE, 2012.

\bibitem{el2013semantic}
Omar El-Gayar and Amit Deokar.
\newblock A semantic service-oriented architecture for distributed model
  management systems.
\newblock {\em Decision Support Systems}, 55(1):374--384, 2013.

\bibitem{han2015network}
Bo~Han, Vijay Gopalakrishnan, Lusheng Ji, and Seungjoon Lee.
\newblock Network function virtualization: Challenges and opportunities for
  innovations.
\newblock {\em IEEE Communications Magazine}, 53(2):90--97, 2015.

\bibitem{hu2015sdn}
Chao Hu, Qun Wang, and Xiuyue Dai.
\newblock Sdn over ip: Enabling internet to provide better qos guarantee.
\newblock In {\em 2015 Ninth International Conference on Frontier of Computer
  Science and Technology}, pages 46--51. IEEE, 2015.

\bibitem{jain2013network}
Raj Jain and Subharthi Paul.
\newblock Network virtualization and software defined networking for cloud
  computing: a survey.
\newblock {\em IEEE Communications Magazine}, 51(11):24--31, 2013.

\bibitem{kim2013improving}
Hyojoon Kim and Nick Feamster.
\newblock Improving network management with software defined networking.
\newblock {\em IEEE Communications Magazine}, 51(2):114--119, 2013.

\bibitem{li2015decentralized}
Juan Li, Nazia Zaman, and Honghui Li.
\newblock A decentralized locality-preserving context-aware service discovery
  framework for internet of things.
\newblock In {\em Services Computing (SCC), 2015 IEEE International Conference
  on}, pages 317--323. IEEE, 2015.

\bibitem{mietz2013semantic}
Richard Mietz, Sven Groppe, Kay R{\"o}mer, and Dennis Pfisterer.
\newblock Semantic models for scalable search in the internet of things.
\newblock {\em Journal of Sensor and Actuator Networks}, 2(2):172--195, 2013.

\bibitem{nunes2014survey}
Bruno Astuto~A Nunes, Marc Mendonca, Xuan-Nam Nguyen, Katia Obraczka, and
  Thierry Turletti.
\newblock A survey of software-defined networking: Past, present, and future of
  programmable networks.
\newblock {\em IEEE Communications Surveys \& Tutorials}, 16(3):1617--1634,
  2014.

\bibitem{saurez2016incremental}
Enrique Saurez, Kirak Hong, Dave Lillethun, Umakishore Ramachandran, and Beate
  Ottenw{\"a}lder.
\newblock Incremental deployment and migration of geo-distributed situation
  awareness applications in the fog.
\newblock In {\em Proceedings of the 10th ACM International Conference on
  Distributed and Event-based Systems}, pages 258--269. ACM, 2016.

\bibitem{seddiki2014flowqos}
M~Said Seddiki, Muhammad Shahbaz, Sean Donovan, Sarthak Grover, Miseon Park,
  Nick Feamster, and Ye-Qiong Song.
\newblock Flowqos: Qos for the rest of us.
\newblock In {\em Proceedings of the third workshop on Hot topics in software
  defined networking}, pages 207--208. ACM, 2014.

\bibitem{sezer2013we}
Sakir Sezer, Sandra Scott-Hayward, Pushpinder~Kaur Chouhan, Barbara Fraser,
  David Lake, Jim Finnegan, Niel Viljoen, Marc Miller, and Navneet Rao.
\newblock Are we ready for sdn? implementation challenges for software-defined
  networks.
\newblock {\em IEEE Communications Magazine}, 51(7):36--43, 2013.

\bibitem{webopedia}
Forrest Stroud.
\newblock Fog computing.
\newblock http://www.webopedia.com/TERM/F/fog-computing.html.

\bibitem{wang2004image}
Zhou Wang, Alan~C Bovik, Hamid~R Sheikh, and Eero~P Simoncelli.
\newblock Image quality assessment: from error visibility to structural
  similarity.
\newblock {\em IEEE transactions on image processing}, 13(4):600--612, 2004.

\end{thebibliography}
\end{document}